\documentclass[sigconf,natbib=false]{acmart}

\author{Stefano Schuppli}
\authornote{Correspondence via schuppli [at] cscs.ch.}
\affiliation{
  \institution{ETH Zurich, Swiss National Supercomputing Centre (CSCS)}
  \city{Lugano}
  \country{Switzerland}
}

\author{Fawzi Mohamed}
\affiliation{
  \institution{ETH Zurich, Swiss National Supercomputing Centre (CSCS)}
  \city{Lugano}
  \country{Switzerland}
}

\author{Henrique Mendonça}
\affiliation{
  \institution{ETH Zurich, Swiss National Supercomputing Centre (CSCS)}
  \city{Lugano}
  \country{Switzerland}
}

\author{Nina Mujkanovic}
\affiliation{
  \institution{ETH Zurich, Swiss National Supercomputing Centre (CSCS)}
  \city{Lugano}
  \country{Switzerland}
}

\author{Elia Palme}
\affiliation{
  \institution{ETH Zurich, Swiss National Supercomputing Centre (CSCS)}
  \city{Lugano}
  \country{Switzerland}
}

\author{Dino Conciatore}
\affiliation{
  \institution{ETH Zurich, Swiss National Supercomputing Centre (CSCS)}
  \city{Lugano}
  \country{Switzerland}
}

\author{Lukas Drescher}
\affiliation{
  \institution{ETH Zurich, Swiss National Supercomputing Centre (CSCS)}
  \city{Lugano}
  \country{Switzerland}
}

\author{Miguel Gila}
\affiliation{
  \institution{ETH Zurich, Swiss National Supercomputing Centre (CSCS)}
  \city{Lugano}
  \country{Switzerland}
}

\author{Pim Witlox}
\affiliation{
  \institution{ETH Zurich, Swiss National Supercomputing Centre (CSCS)}
  \city{Lugano}
  \country{Switzerland}
}

\author{Joost VandeVondele}
\affiliation{
  \institution{ETH Zurich, Swiss National Supercomputing Centre (CSCS)}
  \city{Lugano}
  \country{Switzerland}
}

\author{Maxime Martinasso}
\affiliation{
  \institution{ETH Zurich, Swiss National Supercomputing Centre (CSCS)}
  \city{Lugano}
  \country{Switzerland}
}

\author{Thomas C. Schulthess}
\affiliation{
  \institution{ETH Zurich, Swiss National Supercomputing Centre (CSCS)}
  \city{Lugano}
  \country{Switzerland}
}

\author{Torsten Hoefler}
\affiliation{
  \institution{ETH Zurich, Swiss National Supercomputing Centre (CSCS)}
  \city{Lugano}
  \country{Switzerland}
}

\AtBeginDocument{}

\RequirePackage[
  datamodel=acmdatamodel,
  style=acmnumeric,
  ]{biblatex}

\renewbibmacro*{title}{\printtext{\mkbibemph{\thefield{title}}}}

\setcopyright{acmlicensed}
\copyrightyear{2025}
\acmYear{2025}
\acmDOI{XXXXXXX.XXXXXXX}

\acmConference[CUG25]{Computing Horizons: CUG 2025}{May 04--08, 2025}{New Jersey}
\acmISBN{978-1-4503-XXXX-X/2025/05}

\usepackage{listings}

\definecolor{bgcolor}{gray}{0.95}              

\definecolor{yamlkeys}{rgb}{0.0, 0.0, 0.6}     
\definecolor{yamlstrings}{rgb}{0.58, 0, 0.82}  
\definecolor{yamlcomments}{rgb}{0.5, 0.5, 0.5} 
\lstdefinestyle{yaml}{
    backgroundcolor=\color{bgcolor},           
    commentstyle=\color{yamlcomments},        
    keywordstyle=\color{yamlkeys}\bfseries,   
    stringstyle=\color{yamlstrings},          
    basicstyle=\ttfamily\scriptsize,          
    breakatwhitespace=false,                  
    breaklines=true,                          
    captionpos=b,                             
    keepspaces=true,                          
    numbers=left,                             
    numbersep=5pt,                            
    showspaces=false,                         
    showstringspaces=false,                   
    showtabs=false,                           
    tabsize=4,                                
    morekeywords={true, false, null}          
}

\definecolor{tomlkeys}{rgb}{0.0, 0.0, 0.6}     
\definecolor{tomlstrings}{rgb}{0.58, 0, 0.82}  
\definecolor{tomlnumbers}{rgb}{0.5, 0.1, 0.1}  
\definecolor{tomlcomments}{rgb}{0.5, 0.5, 0.5} 
\lstdefinestyle{toml}{
    backgroundcolor=\color{bgcolor},
    commentstyle=\color{tomlcomments}\itshape,       
    keywordstyle=\color{tomlkeys}\bfseries,          
    stringstyle=\color{tomlstrings},
    basicstyle=\ttfamily\scriptsize,
    breaklines=true,
    captionpos=b,
    keepspaces=true,
    numbers=left,
    numbersep=5pt,
    showspaces=false,
    showstringspaces=false,
    showtabs=false,
    tabsize=4,
    morekeywords={true,false},                       
    morecomment=[l]\#,                               
    morestring=[b]",                                 
    morestring=[b]'                                  
}

\lstdefinestyle{console}{
    backgroundcolor=\color{bgcolor},
    basicstyle=\ttfamily\scriptsize,
    breaklines=true,
    rulecolor=\color{black},
    breaklines=true,
    captionpos=b,
    keepspaces=true,
    numbers=left,
    numbersep=5pt,
    showspaces=false,
    showstringspaces=false,
    showtabs=false,
    tabsize=4,
    keywordstyle=\color{cmd}\bfseries,
    commentstyle=\color{output},
    escapeinside={(*@}{@*)},
}

\usepackage{float}

\usepackage{enumitem}
\usepackage{hyperref}
\usepackage{cleveref}
\crefformat{enumi}{#2#1#3}

\begin{document}

\title{Evolving HPC services to enable ML workloads on HPE Cray EX}

\renewcommand{\shortauthors}{Schuppli et al.}


\begin{abstract}
The Alps Research Infrastructure leverages GH200 technology at scale, featuring 10,752 GPUs. Accessing Alps provides a significant computational advantage for researchers in Artificial Intelligence (AI) and Machine Learning (ML). While Alps serves a broad range of scientific communities, traditional HPC services alone are not sufficient to meet the dynamic needs of the ML
community. This paper presents an initial investigation into extending HPC service capabilities to better support ML workloads. We identify key challenges and gaps we have observed since the early-access phase (2023) of Alps by the Swiss AI community and propose several technological enhancements. These include a user environment designed to facilitate the adoption of HPC for ML workloads, balancing performance with flexibility; a utility for rapid performance screening of ML applications during development; observability capabilities and data products for inspecting ongoing large-scale ML workloads; a utility to simplify the vetting of allocated nodes for compute readiness; a service plane infrastructure to deploy various types of workloads, including support and inference services; and a storage infrastructure tailored to the specific needs of ML workloads. These enhancements aim to facilitate the execution of ML workloads on HPC systems, increase system usability and resilience, and better align with the needs of the ML community. We also discuss our current approach to security aspects. This paper concludes by placing these proposals in the broader context of changes in the communities served by HPC infrastructure like ours.
\end{abstract}

\keywords{HPC, Machine Learning, Research Infrastructure, Platforms}

\maketitle

\section{Introduction}
\label{sec:introduction}

The Alps Research Infrastructure~\cite{Alps}, operated by the Swiss National Supercomputing Centre (CSCS), reached full capacity in 2024 and ranks among the world's top ten supercomputers. With 10,752 NVIDIA Grace-Hopper GPUs (GH200), Slingshot-11 interconnect, and flash-based storage, Alps is instrumental in advancing large-scale ML efforts, particularly as essential infrastructure for the compute and data needs of the "Swiss AI Initiative"~\cite{SwissAIInitiative}.

Alongside the deployment of the Alps infrastructure, CSCS has been developing technologies to enable flexible service offerings for various scientific communities. This technology, known as versatile software-defined clusters or \textit{vClusters}~\cite{vCluster2, vCluster1}, allows the grouping of compute and data resources into clusters or platforms (sets of clusters) and independently deploy services tailored to the needs specific to the communities accesing them. To facilitate the management of multiple platforms, the vCluster technology is designed with descriptive and generic service definitions, enabling automatic deployment of services on the vClusters following pipeline execution. One platform is dedicated to the ML community with its set of specific services. The vCluster technology is instrumental to the flexibility required by the proposals discussed later in this paper.

The ML community, whether academic or commercial, operates within a large, fast-paced, and vibrant ecosystem of tools and services driven by community and vendor initiatives. Consequently, the community's expectations for using an HPC system like Alps go beyond just accessing computational resources through programming environments and batch schedulers. Enabling such an ML ecosystem on HPC service offerings is challenging and underscores the need to evolve HPC services without compromising performance or sustainability.

In this paper, we present an initial investigation into the evolution of HPC services to better serve ML workloads. Based on early user engagement with the Swiss AI community and ongoing internal development efforts at CSCS, we identify gaps and pain points encountered by ML users on Alps. In response, we propose a pragmatic approach based on a suite of modular technological components, including support for containerized user environments, performance profiling tools, infrastructure observability services, node verification tools, a service plane infrastructure to support non-HPC specific workloads, and storage options tailored for ML data requirements. These components are being designed and developed collaboratively with the ML community using Alps, ensuring they effectively address real-world needs and challenges.

The document is organized as follows. Section~\ref{sec:motivations} introduces the tension points and expectation gaps observed since Alps' early-access phase. Section~\ref{sec:tech_components} presents our approach through several proposed technological components. Sections~\ref{sec:discussion}~and~\ref{sec:conclusions} conclude by contextualizing the motivations and proposed approaches within the overall shift of interests that centers like ours are currently experiencing.


\section{Motivations}
\label{sec:motivations}

Running ML workloads on HPC infrastructure presents several challenges that must be addressed to support adoption and enhance ML user productivity, while maintaining a focus on critical aspects such as computational efficiency. These challenges can be summarized as follows:

\begin{enumerate}[label=\textbf{(\alph*)}]
    \item \label{item:problem_stmt:hpc_kb_gap} \textbf{HPC knowledge gap}:
    The rise of large ML models has led many in the ML community to adopt HPC systems. However, new users often encounter adoption barriers due to unfamiliar components, such as queue-based workload managers, and may adopt suboptimal approaches to meet their needs. While the basics can be learned, improper usage can negatively impact other users. Effective performance engineering, essential for optimal resource utilization, demands significant expertise and time. Users lacking this expertise may struggle to fully leverage the expensive underlying resources entrusted to them.

    \item \label{item:problem_stmt:support_diverse_and_evolving_needs} \textbf{Diverse and evolving needs:} 
     While HPC systems are well-suited for large-scale ML training, typical ML projects and workflows encompass a broader range of requirements and stages. These include ancillary supporting services for model development and exploration, dataset acquisition and preparation, and inference-oriented workload managers for validation and model deployment. HPC job schedulers are not designed to support this breadth of tasks, often leading to inefficient resource utilization when used for such purposes. Addressing the rapidly evolving requirements of ML workflows necessitates more flexible and adaptive approaches and the appropriate set of services.
        
    \item \label{item:problem_stmt:rework_and_productivity} \textbf{Repetitive work affecting productivity:} Common ML challenges such as detecting inefficiencies, handling infrastructure-caused interruptions, or debugging throughput variability sources, often lead teams to adopt localized ad-hoc strategies, resulting in redundant efforts between teams and missed opportunities to improve the infrastructure for the benefit of all users.
    
    \item \label{item:problem_stmt:reproducibility_and_portability} \textbf{Reproducibility and portability:} ML workloads need to be replicable across different infrastructures to provide users with the flexibility they need across different project stages, and to prevent lock-in situations.

    \item \label{item:problem_stmt:storage_offering} \textbf{Storage offerings alignment:} ML workloads can present different storage access patterns, not only across project phases but also within the same application execution. Conventional storage solutions like parallel file systems may not meet all these needs, sometimes necessitating application-level workarounds to maintain efficient I/O operations. Tailoring storage services to better serve ML workloads can facilitate users and at the same time allow investments refinement, e.g., by distinguishing beneficial storage features from superfluous ones.
    
    \item \label{item:problem_stmt:ops_support_constraints} \textbf{Operational and support resource constraints:} Publicly funded HPC centers operate with resource constraints that are more limiting than those of leading commercial centers. This makes it more difficult to develop and maintain comprehensive solutions while ensuring system reliability.

    \item \label{item:problem_stmt:security_considerations} \textbf{Security considerations:} As a public HPC provider, security and ethical concerns are longstanding priorities. However, the integration of ML workloads introduces novel challenges, including specific confidentiality and privacy needs, and potential misuse of the infrastructure.

\end{enumerate}

Addressing these challenges is crucial to enabling productive ML research on HPC infrastructure.

\vspace{1.4cm}

\section{Technological components}
\label{sec:tech_components}

\subsection{Support for Container-based User Environments}
\label{subsec:containers_support}
\subsubsection{Motivation}
While novel approaches to managing HPC software stacks, such as \texttt{uenv}~\cite{uenv} (based on \texttt{Spack}~\cite{spack}), are also relevant to the ML community, ML users are typically already familiar with container-based workflows. They are also already aware of vendor-curated, ready-to-use containers\footnote{Prominent examples of vendor-curated container catalogs are the NVIDIA NGC Catalog \url{https://catalog.ngc.nvidia.com} and the AMD Infinity Hub \url{https://www.amd.com/en/developer/resources/infinity-hub.html}.} for established libraries such as PyTorch and JAX. Containers also provide users with more control over their own user space, thus facilitating the installation of custom dependencies. This increased autonomy is crucial in the fast-evolving ML field. Python dependencies not available in base images can be included rapidly through bind-mounted virtual environments. ML users appreciate an experience that closely mirrors their familiar environments, enabling a relatively quick start by minimizing the upfront HPC-specific knowledge required to use the infrastructure. By anticipating future project stages, users also value the portability of container technologies for deploying models across different platforms.

\subsubsection{Proposed approach}
Since 2016, CSCS has invested in supporting container technologies for HPC workloads~\cite{sarus}, leading to the current comprehensive container-support offering. This includes a toolset, named Container Engine (CE)~\cite{containers-first-ce}, designed to enable computing jobs to seamlessly run inside Linux application containers, along with a set of Open Container Initiative (OCI)~\cite{oci} hooks and soon Container Device Interface (CDI)~\cite{cdi} specifications designed and maintained by CSCS to transparently optimize performance-critical operations for users. OCI hooks are typically used to inject files (e.g., libraries) or set configurations at container creation to achieve native performance for storage, GPU compute and network.

\vspace{.5cm}
\begin{lstlisting}[style=toml, caption=Example of an Environment Definition File (EDF)., label=sec:containers:edf-content-example]
image = "ubuntu:latest"

mounts = [
    "/scratch/project01/dataset:/scratch/project01/dataset:ro",
    "/scratch/${USER}"
]

workdir = "/scratch/${USER}/project01_code"
writable = true

[annotations.com.hooks.aws_ofi_nccl]
variant = "cuda12"

[annotations.com.hooks.ssh]
enabled = "true"
authorize_ssh_key = "<public key file path>"
\end{lstlisting}

\subsubsection{Design elements}
An important design point is to provide users with the ability to define their environment in a clear and concise way.
This was achieved using a TOML-based Environment Definition File (EDF) shown in
Listing~\ref{sec:containers:edf-content-example}. Its syntax should appear familiar to users who have already experienced using containers. Line 12 exemplifies a behavior customization for a default performance-focused OCI hook. Lines 14-16 illustrate exposing an SSH port without the need to build a new container image (e.g., to install OpenSSH), useful for rapidly setting-up IDE-based debugging sessions on the same software environment on which the applications are going to run. An Environment Definition File can then be used as shown in Listing~\ref{sec:containers:edf-usage-example}, Line 6 which executes the command inside the specified EDF environment.

\begin{lstlisting}[style=console, caption=Example of EDF files usage with Slurm., label=sec:containers:edf-usage-example]
#!/bin/bash

#SBATCH --nodes 64
#SBATCH --ntasks-per-node 4

srun --environment=./my_environment.toml python train.py
\end{lstlisting}

Additional components include pre-tuned system configurations (e.g., for NCCL or RCCL), and integrations to simplify common tasks like IDE-based debugging and profiling within the same software environment used for jobs execution.

\subsubsection{Ongoing work and future directions} 
While the use of EDF abstracts the use of container tools at runtime, we are working on integrating the image building process, as well as improving the integration with vulnerability scanning services usually found in image registries. We would also like to provide users with the possibility to define start-up steps, which could be helpful to support common patterns such as the activation of Python virtual environments. Furthermore, we are in the process of transitioning to Podman as the central component of our tool set.

To further assist users, we also aim to provide documentation with quick-start blueprints, performance baselines, checklists (e.g., Alps-specific NCCL or RCCL tuning configurations), and technical reports (e.g., on dataloader options relevant for containerized contexts). Lastly, user-contributed performance regression monitoring enhances system reliability. While our aim is to make user interfaces easier to adopt, broad container support also enhances the ML software stack portability, thus facilitating MLOps-inspired workflows involving  models deployment to Kubernetes clusters when needed.

\subsubsection{Relevance} Users can define software environments in familiar formats and specify them at job submission, streamlining interaction with the system, and easing application portability across platforms. As such, this proposal contributes to addressing challenges \cref{item:problem_stmt:hpc_kb_gap}, \cref{item:problem_stmt:support_diverse_and_evolving_needs} and \cref{item:problem_stmt:reproducibility_and_portability} as introduced in Section~\ref{sec:motivations}.

\subsection{A GPU Saturation Scorer for ML Applications}
\label{subsec:gpu_saturation_scorer}
\subsubsection{Motivation}
In the last decade, GPUs have become a cornerstone in the TOP500 list. In 2011, the second fastest and three of the top ten supercomputers were GPU-powered~\cite{mcintosh2011gpu}. By 2024, 41.8\% of all machines were accelerated systems~\cite{amdnvidiatop500}. Modern GPUs have complex architectures offering exceptional performance and high energy efficiency for large, parallelized, high-throughput workloads. GPUs are thus ideal for ML workloads, but utilizing their full potential requires careful calibration. Interpreting the numerous performance metrics might be overwhelming for users that lack computer science fundamentals.

A common complaint from our users involves performance variability in workloads that, nevertheless, display constant GPU utilization. Seemingly minor code changes such as, e.g., the usage of alternative fused operation implementations, may lead to improved performance. Such potential might not be evident to inexperienced users, especially if stable high GPU utilization is observed. The GPU utilization metric, commonly obtained via \texttt{nvidia-smi} or through the default configuration of \textit{Weights \& Biases}, itself NVML-based, could mislead novice users. Understanding the distinction between GPU utilization as the average time a resource was accessed and its saturation as the degree to which the resource was loaded requires a deeper analysis beyond surface-level metrics~\cite{understanding_nvidia_GPU_utilization_vs_saturation}. This difference in understanding of utilization as a temporal metric, and saturation as the actual spatial metric, may be confusing to users who equate utilization with resource loading. 

NVIDIA provides various proprietary profiling and analysis tools for their GPUs, targeting different levels of granularity. These include comprehensive tools such as the NVIDIA Nsight suite, consisting of Nsight Compute (\texttt{ncu}) and Nsight System (\texttt{nsys}). \texttt{ncu}, designed for collecting fine-grained, instruction-level performance data for individual kernels or small sets of kernels, provides the higher detail of the two applications. This level of detail can be problematic, as it leads to significant overhead as well as memory usage when evaluating large, distributed workloads end-to-end. \texttt{nsys} offers more flexibility by combining sampling and tracing capabilities to capture both fine-grained and coarse-grained profiling information at the system level, making it more compatible with distributed workloads, but also prohibitively expensive for large ones. Effectively interpreting the results of either tool often demands significant time and technical expertise, which users may lack. 

\subsubsection{Proposed approach}
To address the challenge of GPU activity metrics not reflecting performance differences, we propose a \textit{GPU saturation scorer} utility that returns a simple assessment score for efficiency evaluation, as well as providing options for obtaining greater workload insights. The tool is built to leverage the data produced by underlying systems, such as the NVIDIA Data Center GPU Manager (DCGM), and be integrated with the Cray EX telemetry system.

DCGM is a tool for fine-grained and targeted metric gathering that operates as a lightweight daemon with minimal overhead, making it suitable for continuous use. While the daemon process requires root privileges, non-privileged users can interact with it through its API to access metrics data. DCGM also supports telemetry, enabling continuous metric gathering and storage in e.g. the Cray EX telemetry system database for later querying. 

Limitations of DCGM include a lack of native support for isolating resources allocated to specific workloads by the scheduler, thus focusing on cluster-level data querying instead of on the node group level. This limitation aside, our analysis of different possible approaches identified DCGM as the most suitable base tool candidate for our light-weight, user-friendly, and privilege-safe solution for assessing the efficient use of hardware for large distributed GPU workloads.

\begin{table*}
    \begin{tabular}{ll}
        \toprule
        Performance Metric&Description\\ 
        \midrule
        Graphics Engine Activity&Fraction of time any portion of the graphics or compute engines was active.\\ 
        SM Activity&Fraction of time at least one warp was active on a multiprocessor.\\ 
        SM Occupancy&Fraction of resident warps on an SM relative to the maximum warps supported.\\ 
        FP Engine Activity&Fraction of cycles the Tensor core/FP64/FP32/FP16 pipe was active.\\ 
        Memory Bandwidth Utilization&Fraction of cycles during which data was sent to or received from device memory.\\ 
        Transfer Bandwidth&Rate of data transmission/reception over PCIe/NVLink.\\
    \bottomrule
\end{tabular}
\caption{A selection of the possible DCGM-accessible performance metrics, that our proposed GPU saturation scorer tool~\cite{DCMG_metrics} aims to make more digestible for end users.}
\label{tab:gpu_metrics}
\end{table*}

\subsubsection{Design elements}
While our GPU saturation scorer tool can be invoked on a single process from the CLI, it is thought for multi-node assessments and it is therefore integrated with Slurm. Each task within a Slurm job step wraps its workload process by invoking the saturation scorer. Information collected by the utility includes (1) the nodes involved in the workload, (2) the number of tasks per node, (3) the number of GPUs allocated per task, and (4) the specific GPUs associated with each process. With this information, the scorer connects to the local DCGM daemon on each node. It then creates a unique GPU group that contains only the GPUs associated with the process, making it possible to aggregate data across all nodes involved in the distributed workload. 

The raw data collected can then be processed and analyzed to derive a meaningful metric for the user. 

\begin{figure}[ht]
    \centering
    \includegraphics[width=1\linewidth]{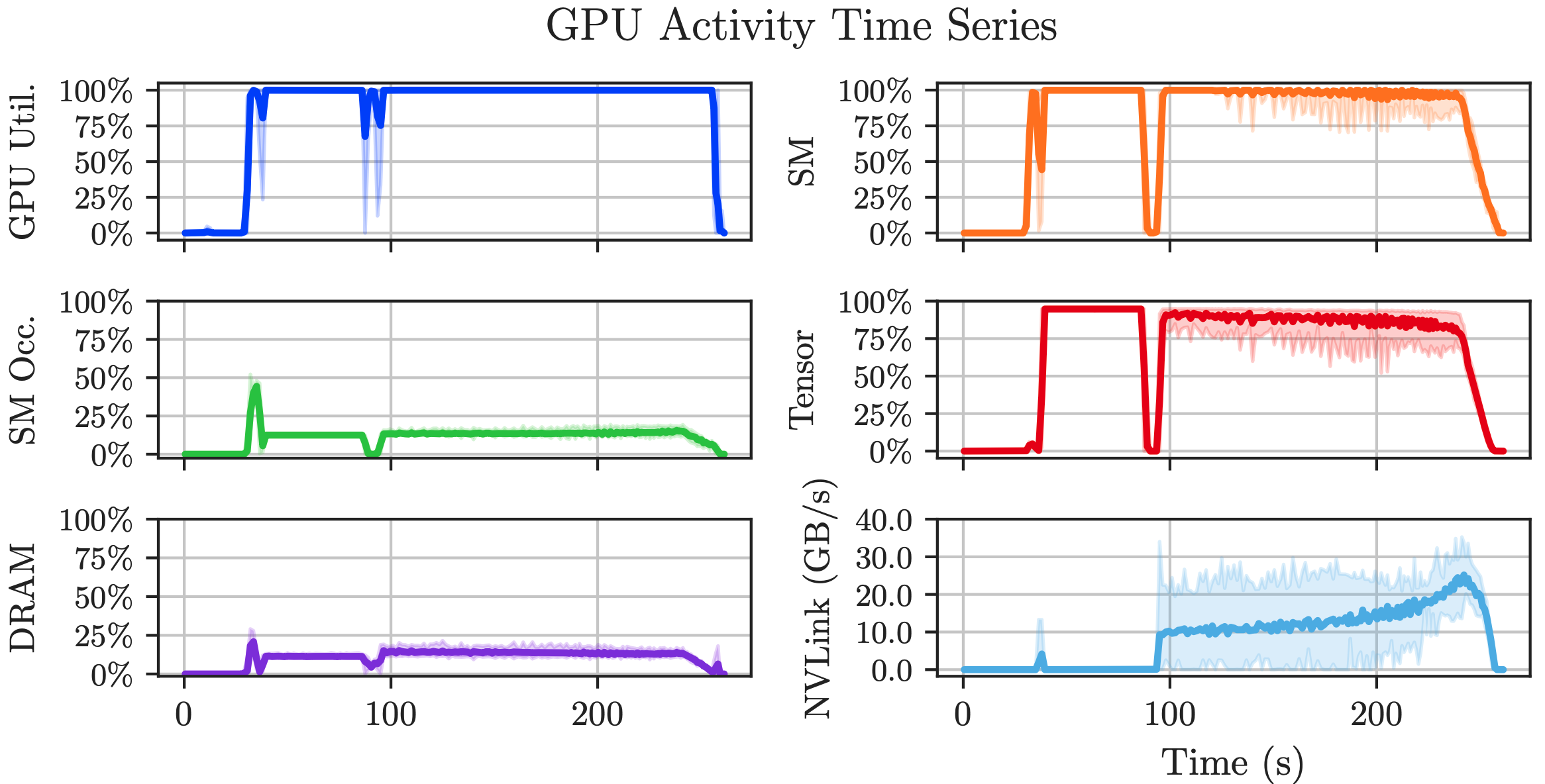}
    \caption{Example GPU activity time series plots generated using output of the GPU saturation scorer tool over a 4 nodes (16 GPUs) benchmark. The first plot visualizes the classic GPU utilization metric; the remaining plots correspond to a number of DCGM performance metrics including SM Occupancy, Memory BW Utilization, SM Activity, Tensor Activity, and NVLink Bandwidth~\cite{hpc_ai_advisory_ferrari_mujkanovic}.}
    \label{sec:HPL_saturation_time_series}
    \Description{Saturation scorer activity time series for 16GPU/4 node HPL run.}
\end{figure}

\begin{figure}[ht]
    \centering
    \includegraphics[width=1\linewidth]{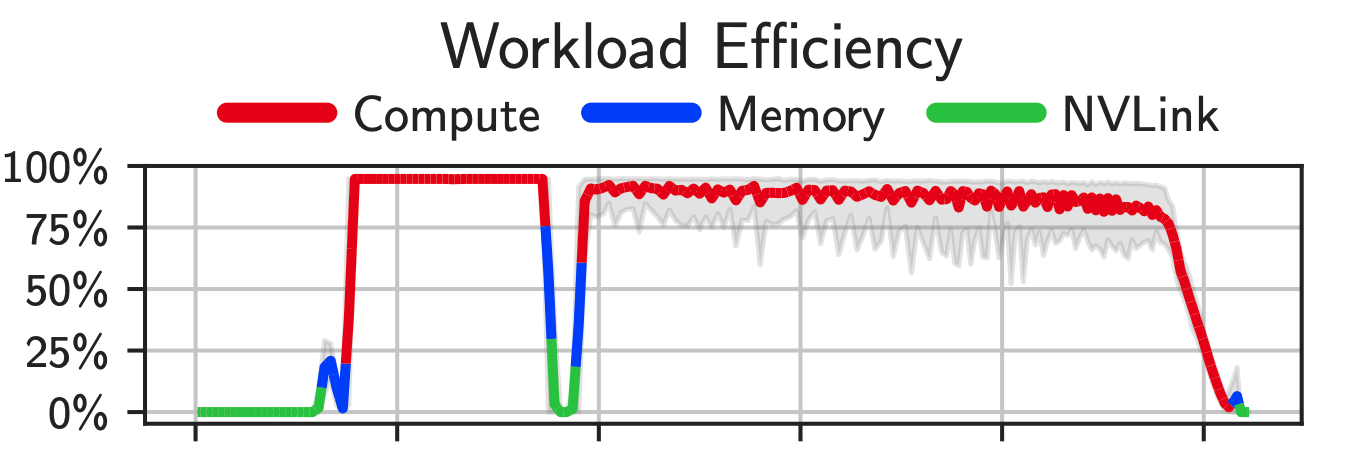}
    \caption{The previous GPU Activity Time 
    Series plots can be aggregated into a single plot for a simplified overview of the impact of compute, memory, and network on the overall GPU saturation score. This plot is a generalized output of a performance model developed for our GPU saturation scorer. Placing more weight on memory or compute data may produce differing plots. Performance modeling is further discussed by Ferrari et. al~\cite{hpc_ai_advisory_ferrari_mujkanovic}. }
    \label{sec:saturation_scorer_aggregate}
    \Description{Saturation Scorer output aggregate.}
\end{figure}

The huge number of GPU metrics available depends on the model, on the hardware parameters, on the resource usage, and the type of GPU activity. We believe the fundamental contribution of our proposed approach stands in the meaningful aggregation of carefully selected metrics. Thus, we analyze selected profiling metrics relevant to users and introduce easy-to-digest performance indicators for supporting their evaluations. An illustrative selection of the performance metrics underlying our saturation score model are listed in Table~\ref{tab:gpu_metrics}. Note that users can manually select additional metrics to monitor. All of these metrics can then further be visualized as time-series plots, illustrating GPU activity changes over time as shown in Figure ~\ref{sec:HPL_saturation_time_series}, and aggregated, illustrating the impact of each performance metric on the overall GPU Utilization metric, as see in Figure~\ref{sec:saturation_scorer_aggregate}.  

We provide this tool to users as a low-overhead, non-intrusive method to attain an initial point of reference for understanding application efficiency. It is designed to complement advanced profiling tools by offering a hardware-specific perspective that goes beyond surrogate indicators such as time per iteration or tokens per second, thus aiding users in gaining a more hardware-specific understanding before delving into detailed analyses with advanced profiling tools. 

\subsubsection{Ongoing work and future directions.}
Future improvements to the GPU saturation scorer include extending the tool to cover additional metrics regarding the network, NCCL and RCCL, and MPI, as well as storage.  A performance model should be derived from these metrics and serve to extend the available visualizations. Further visualization projects involve the mapping of the metrics into heatmaps of GPU activity, as well as the integration into a web-based GUI easily accessible to users.  

Integration of our proposed tool into the Alps container-first environment hinges on its integration with the container tools and Slurm. Closer integration with the Cray EX telemetry system as a central metric collection hub allows for utilizing existing DCGM metrics, facilitating further development and integration of the GPU scorer.

Due to operational needs and the phased installation and gradual availability of different GPU models on Alps, the tool was initially developed to support NVIDIA GPUs using the NVIDIA proprietary DCGM tool. We are now working to extend its capabilities by integrating support for AMD GPUs using their equivalent tools and performance metrics.

\subsubsection{Relevance} This proposal is particularly relevant to challenges \cref{item:problem_stmt:hpc_kb_gap} and \cref{item:problem_stmt:rework_and_productivity} as introduced in Section~\ref{sec:motivations}.

\subsection{Infrastructure Observability for ML Workloads}
\label{subsec:observability}
\subsubsection{Motivation}
As distributed ML workloads scale across increasing numbers of nodes and GPUs, the likelihood of inefficiencies, such as stragglers, resource under-utilization, or suboptimal communication patterns, rises significantly. These issues may arise from a broad range of typically transient factors, such as subtle node health degradation, which can impact overall training runtime even in the absence of explicit errors. 
The task of detecting and then identifying the source of such inefficiencies in a distributed context can be challenging and time-consuming and in the presence of other project priorities, it tends to be neglected. This is especially the case in situations where the performance impact is not evident or acute enough to demand immediate attention.

\subsubsection{Proposed approach}
We aim to assist users and operations teams by (1) offering continuous visibility into running workloads to facilitate the identification of issues, and (2) providing tools to (more rapidly than today) identify the root causes of anomalous workload behaviors, such as inconsistent throughput.

Although for (1) solutions such as \textit{Weight \& Biases}\footnote{\url{http://wandb.ai}} exist and are established in the community, they are limited by the restricted set of metrics collected, by service rate-limits that push users to observe only a small number of processes (e.g., typically rank 0 or ranks 0-3 alone), or by non-parameterized sampling frequencies. As for (2), the opportunity we see is in the systematic organization of system-level debugging expertise. This specialized knowledge generally manifests itself through small-scale debugging logic developed ad-hoc in response to pressing issues,  such as for verifying the presence of degraded network equipment. By providing the facility to capture this logic and maintaining it over time and possibly even running continuously, we aim not only at making it available more rapidly when needed again, but also to surface this expert knowledge to end-users to e.g., increase their autonomy.

By analyzing dependency graphs and communication patterns, the proposed data products empower users to detect optimization opportunities, such as improving resource utilization and pinpointing stragglers processes that may be slowing down collective operations due to poor hardware health status or other imbalances. With these insights, users can, more often than not, independently identify potential enhancements to their workloads and take corrective actions to improve performance and efficiency.

\subsubsection{Design elements}
Our observability stack is built upon CSCS’s Extensible Monitoring and Observability Infrastructure (EMOI)~\cite{emoi}, which ingests telemetry data from Alps into a scalable Elasticsearch backend. Data access is enabled via Kibana, Grafana, programmatic interfaces, and also through case-specific in-house developed web UIs. This infrastructure allows for the aggregation of heterogeneous metrics, from GPU health (such as, temperature throttling, ECC errors, and similar), to Slingshot interconnect counters, including Lustre performance data, in order to feed coherent data products (e.g., visualizations, rapid correlation exploration tools). Key design goals include:
\begin{enumerate}
    \item Job-scoped data products such as dashboards that provide a vertical, per-job view across all ranks, GPUs, and nodes.
    \item Global overlays that situate a job’s performance in the context of the overall Alps system load.
    \item Support for augmentation with optional user-provided metrics such as tokens/sec or iteration latency, to enable further correlation possibilities.
    \item Progressive opt-in to data analysis features to let users learn about more advanced diagnostics possibilities (e.g., allocated nodes fragmentation on the network topology, NCCL performance outliers) in a gradual manner.
\end{enumerate}

\begin{figure}[ht]
    \centering
    \includegraphics[width=\linewidth]{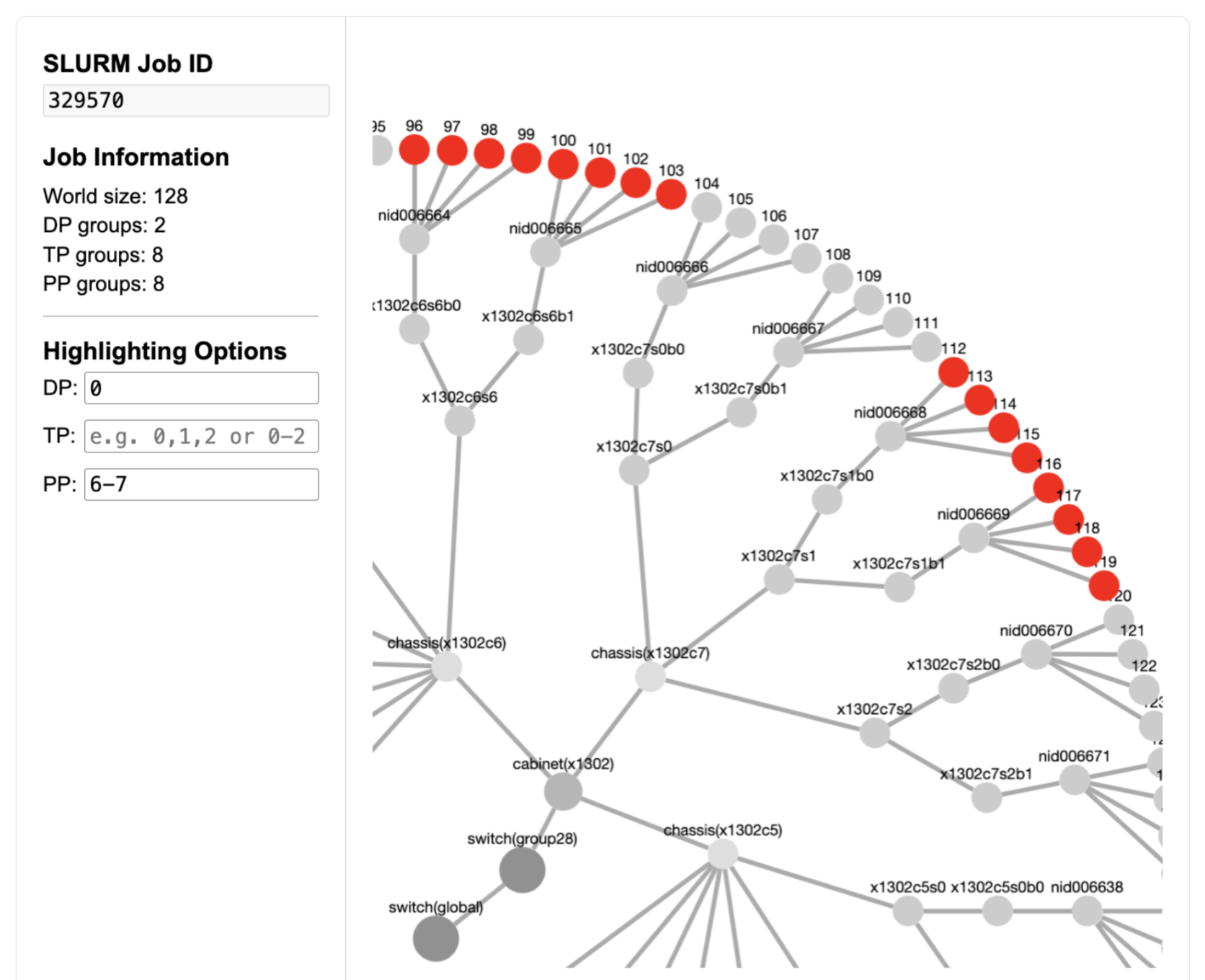}
    \caption{The proof of concept of a web interface meant for inspecting the hardware involved in a given Slurm job. By providing the necessary data, users gain access to additional visualization and filtering features. In this example, the 7\textsuperscript{th} and 8\textsuperscript{th} pipeline layers of the 1\textsuperscript{st} data-parallel model instance in a Megatron-LM training job are highlighted in red, allowing visual analysis of their placement across network components.}
    \label{sec:observability:3d-parallelism-visulization}
    \Description{A diagram of a large tree graph.}
\end{figure}

While collecting user-supplied application metrics (e.g., tokens/sec or iteration latency) may seem redundant with external tools like \textit{Weights \& Biases}, integrating them into the same data infrastructure provides the advantage of enabling correlation and pattern analysis across system layers. Additionally, certain information, such as 3D-parallelism ranks placement over the network topology, are only available within the application context and can only be collected if explicitly exposed by the user. To reduce the user burden in learning about the CSCS service specifics and also in instrumenting their applications, our data products are designed to provide value out of the box. Additional features are enabled via progressive opt-in, by providing the necessary additional data as needed. Pointers to the relevant technical information are provided on the same interface to spread the adoption burden.

The establishment of a catalog of standardized and well documented quality datasets and dataflows further facilitates the development of additional and custom data products also by end-user or operations teams not directly involved with the setup and operations of the underlying data infrastructure.

\subsubsection{Ongoing work and future directions}
Figure~\ref{sec:observability:3d-parallelism-visulization} illustrates a proof of concept where gray points represent already available data (in this case, Slurm job records and information on network components). Additional features, such as filtering options and colored highlighting, become dynamically available once the necessary data is provided by the user. The upfront learning and instrumentation effort can thus be spread and adapted to the needs and interests of the moment.

Similar PoCs are being designed to support use-cases such as *CCL timeouts debugging (leveraging data produced by the PyTorch Flight Recorder \cite{pt_flight_recorder}) and general stragglers detection.

\subsubsection{Relevance}
The proposed solution improves the possibilities users have to autonomously become aware of silent inefficiencies and optimization opportunities. This is done by emphasizing easier access and usability of infrastructure data related to running workloads. This empowers users to detect and act on issues that might otherwise go unnoticed due to project time or expertise limits. Additionally, it preserves and operationalizes expert-developed system-level debugging practices, ensuring they remain accessible and actionable when needed again. As such, this proposal contributes to addressing challenges \cref{item:problem_stmt:hpc_kb_gap}, \cref{item:problem_stmt:rework_and_productivity} and \cref{item:problem_stmt:ops_support_constraints} as introduced in Section~\ref{sec:motivations}.

\subsection{A Node Vetting and Early Abort System}
\label{subsec:nodes_vetting}
\subsubsection{Motivation}
The reliability of allocated nodes is a concern in large-scale ML workloads. The presence within an allocation of a single unhealthy node might cap the performance of the healthy nodes or even prevent the run from completing successfully altogether. The larger the allocation, the higher the likelihood of encountering ``that" unhealthy node. 

HPC centers are interested in making user applications more efficient, but frequent unplanned job interruptions, even in the presence of frequent model checkpointing~\cite{reliability_in_largescale_ML}, can void the benefits produced by performance engineering efforts.

Ensuring the health of the allocated nodes before execution is crucial to improving overall cluster productivity. Numerous recent publications on large ML models include sections dedicated to their training operational experiences~\cite{llama3_interruptions, geminiteam2024, olmo2furious}.

Faced with repetitive system-caused interruptions, users might explicitly exclude unreliable nodes, e.g., via Slurm \verb|--exclude|. This should be avoided. It is our goal as HPC system operators to provide reliability, but new systems might take time to stabilize. Meanwhile, large-scale ML training users need a solution ``today".

\subsubsection{Proposed solution}
To meet this need, we propose the {\emph{Nodes Vetting and Early Abort System}}, a dynamic solution that helps users verify node readiness with rapid, lightweight diagnostic tests just before their application execution. In contrast to regression or integration tests regularly executed at system level, the node vetting tests are meant to catch more dynamic issues such as high GPU temperature, low available memory or other ``dirty" GPU states, network congestion, and similar. Furthermore the set of tests to execute is not meant to be exhaustive but rather focusing on catching the most frequent offending nodes.

This solution is complementary to Slurm's prologue and epilogue and its usage is optional. It is intended for jobs involving significant amounts of resources, thus leaving smaller-scale, iterative development activities unaffected from larger overheads that only amortize on large scale runs. 

\begin{figure}[ht]
     \centering
     \includegraphics[width=\linewidth]{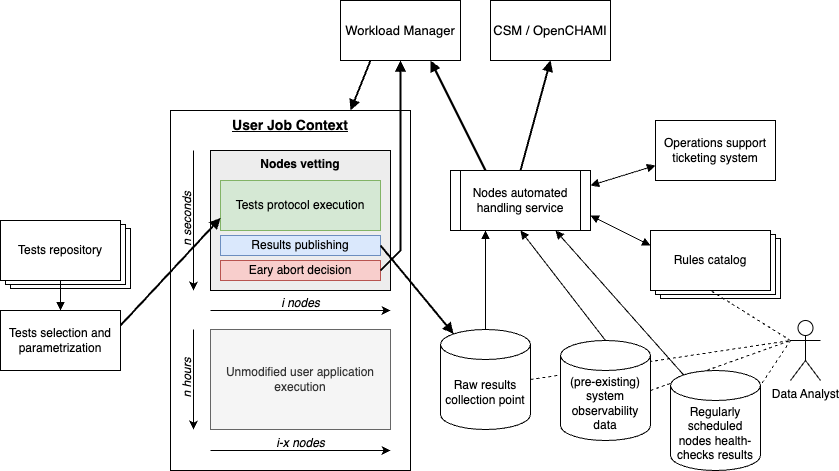}
     \caption{Components overview for the nodes vetting and early abort system.}
     \label{fig:vetting}
     \Description{A diagram that shows the node vetting components and their relations.}
 \end{figure}

Figure~\ref{fig:vetting} shows the components involved: a repository of tests, a CLI tool utility for the user, a data collection service, a catalog of rules to interpret tests results, and an automated nodes handling service.

The tests are organized in a community-contributed repository. Users select those relevant for their workload, composing a {\emph{tests protocol}}. This comes in the form of a yaml configuration file, listing the selected tests to be executed and the tolerated test outcome values as shown in Listing~\ref{listing:vetting}. Tests requirements are also listed as part of the tests protocol and will be dynamically installed prior to the tests execution.

The resulting test protocol is then executed by means of a Python CLI tool as part of the job script, just before the application execution. The test outcome will inform if the entire job should be aborted or if the execution can continue.
For mixed outcome (e.g., the presence of a GPU significantly hotter than the rest) if the main workload allows for a flexible node allocation the unhealthy nodes might be excluded from the job next steps, so to avoid the need of re-queuing.

Additionally, the test outcome can be (w.r.t., user opt-in) collected in a central storage for the HPC system operators to consult and act upon.
Centralizing such node reliability information provides a shared knowledge to simplify operational efficiency, to the benefit of all system users. As part of the shared back-end, a service observes incoming test results and takes action. For example, for repetitive offending nodes, it might exclude them from allocatable resources and take incremental recovery steps and open support tickets as a final fallback.

Comparable approaches within their respective organizations are mentioned in~\cite{megascale, a_practitioner_guide, from_bare_metal_to_a_70B_model}.

\begin{lstlisting}[style=toml, caption=Example of Node Vetting Protocol, label=listing:vetting]
name: "ML Training Node Vetting"
evals:
- name: "Check GPU"
  type: vetnode.evaluations.gpu_eval.GPUEval
  max_temp:  30 #(celsius)
  max_used_memory: 0.2 #(%)
- name: "NCCLBandwidth"
  type: vetnode.evaluations.nccl_eval.NCCLEval
  min_bandwidth: 90.0 #(GBps)
  requirements:
    - torch
- name: "CudaKernel"
  type: vetnode.evaluations.cuda_eval.CUDAEval
  requirements:
    - cuda-python
    - numpy
\end{lstlisting}

\subsubsection{Relevance}
This solution standardizes nodes vetting, offloading project teams from devising strategies to cope with nodes verification. It unifies efforts across teams, such as test definitions and operational information sharing and address challenges \cref{item:problem_stmt:hpc_kb_gap}, \cref{item:problem_stmt:rework_and_productivity} and \cref{item:problem_stmt:ops_support_constraints} as introduced in Section~\ref{sec:motivations}.

\subsection{Service Plane for Supporting and Inference Services}
\label{subsec:service_plane}
\subsubsection{Motivation}
We refer to \textit{supporting services} as deployable services meant to facilitate teams in their project activities. Examples of these are experiment tracking products and workflow engines. Such services are typically lightweight (i.e., commodity hardware suffices), run continuously, and require data persistence. Such products are often available through SaaS offerings, but this is not always the case (e.g., MLFlow), nor suitable for all situations (e.g., confidentiality and particular rate-limit needs).

The absence of a solution for deploying supporting services close to Alps might force users to allocate high-end nodes for makeshift solutions to have such services running alongside their training allocations. It is impractical for HPC centres like ours to operate these services on behalf of users due to resource limits and diverse community needs. Also, choosing among established products is made difficult by the different teams' preferences.

\subsubsection{Proposed approach}
We propose the introduction of a dedicated infrastructure that empowers users to deploy and manage services independently, supported by a community-driven catalog of blue-prints to facilitate adoption.

Additionally, extending this service plane to support inference workloads broadens its utility, enabling additional use cases.

Using Kubernetes for this purpose enables fault-tolerant inference services, unifies commodity hardware with high-end GPU nodes under a single service plane, and leverages existing containerized options on Slurm (w.r.t., Section~\ref{subsec:containers_support}) to facilitate MLOps-inspired training and deployment workflows. Furthermore, its ubiquity simplifies the portability of inference workloads to infrastructures beyond CSCS.

Alternatives, such as Cloud-based one, for deploying inference and supporting services are feasible, especially with tools like FirecREST~\cite{FirecREST} enabling
programmatic access to HPC resources via restful protocols. However, their viability may be constrained by factors such as cost, the need for direct access (e.g., for data handling), or specific use case demands (e.g., frequent large models movements).

\begin{figure}[ht]
    \centering
    \includegraphics[width=\linewidth]{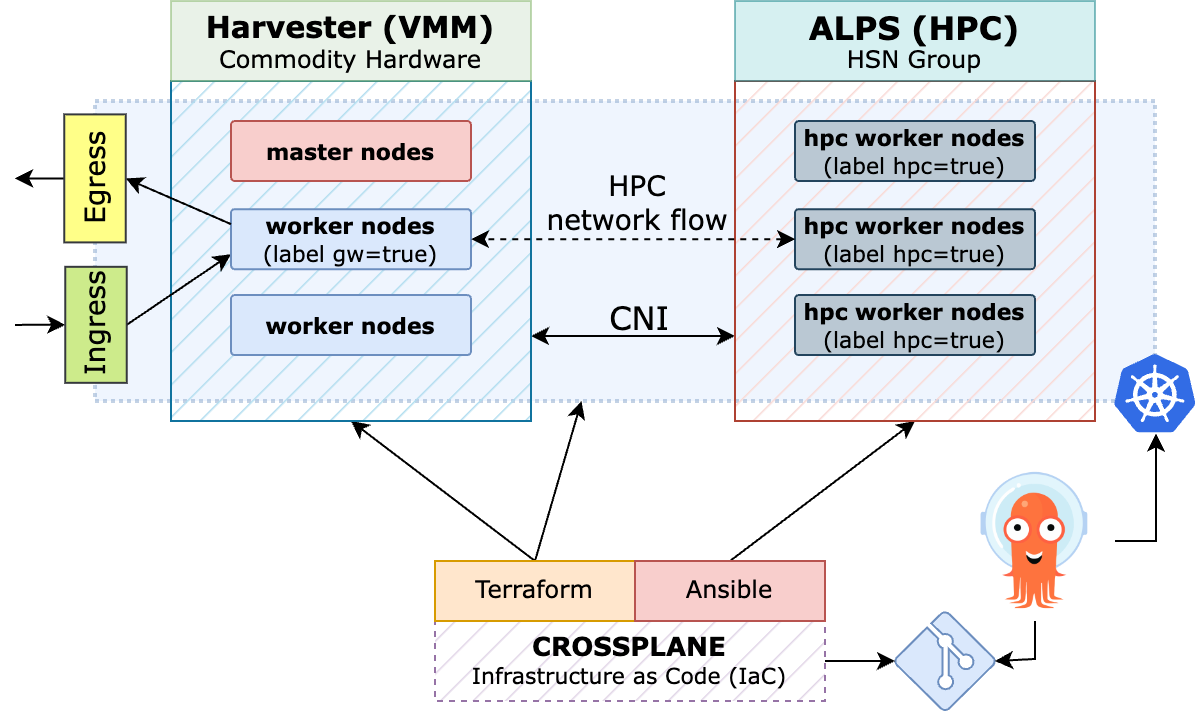}
    \caption{Hybrid Kubernetes service plane integrating virtualized infrastructure (Harvester) and HPC nodes from Alps via Cilium-based CNI. Control plane and lightweight services run on VMs, while GPU workloads execute on HPC nodes. Infrastructure is managed through Crossplane with Terraform and Ansible, enabling reproducible deployments and user-controlled service instantiation.}
    \label{fig:alpernetes}
    \Description{A diagram showing the relationship between Harvester and Alps.}
\end{figure}

\subsubsection{Design elements} The envisioned solution is composed by the following elements, also illustrated in Figure~\ref{fig:alpernetes}:

\begin{itemize}
    \item \textbf{RKE2\footnote{RKE2 is Rancher's enterprise-ready next-generation Kubernetes distribution. See also \url{https://docs.rke2.io}.}-based architecture}. The envisioned architecture integrates a Kubernetes cluster using RKE2 across virtual machines (VMs) based on commodity hardware and HPC nodes from Alps. The control plane and low-resource services run on VMs, which serve as master nodes hosting components like the API server, DNS, and ingress controllers. This efficient setup minimizes load on HPC resources. HPC nodes are dedicated to GPU-intensive tasks and are hidden from the Internet for security reasons, using Cilium for network traffic management. Floating IPs enable external communication, ensuring HPC resources focus on compute workloads while remaining integrated with Kubernetes.
  
    \item \textbf{Automated deployment with ArgoCD for GitOps}. To maintain consistency, traceability, and automation in cluster and applications deployment, ArgoCD is used for GitOps-oriented operations. All infrastructure and application changes are defined declaratively and stored in Git repositories, ensuring a single source of truth and enabling auditable, reproducible deployments with easy rollbacks for change tracking and disaster recovery purposes.
    Rancher's project-based multi-tenancy enhances namespace-level isolation. Each team or workload gets a dedicated namespace within a project, enforcing boundaries and allowing independent team operations.
    ArgoCD's \texttt{ApplicationSets} deploy and manage common services across clusters or namespaces. With templated configurations and dynamic generators, \texttt{ApplicationSets} simplify applications replication across environments, minimizing configuration drift and ensuring uniform updates. This model scales well with cluster growth and improves manageability of distributed Kubernetes environments.

    \item \textbf{GPU Operators}. GPU workload orchestration in Kubernetes on HPC nodes requires a dedicated layer for abstraction and configuration of GPU resources. This involves labeling GPU-capable nodes, defining Custom Resource Definitions (CRDs), and installing necessary drivers and runtime components for effective resources usage. We use NVIDIA and AMD GPU Operators to automate management tasks such as driver installation and runtime configuration, reducing operational complexity. These operators are deployed only on HPC nodes to handle intensive workloads, preserving VMs for control and services.

    \item \textbf{Flexible storage strategy}. The hybrid Kubernetes architecture uses flexible storage options to support various workloads, from temporary compute jobs to persistent applications. Kubernetes \texttt{StorageClasses} abstract storage implementation, enabling dynamic volume provisioning based on specific workload needs. Longhorn manages local storage for critical high-performance or node-locality tasks by providing a lightweight, cloud-native block storage solution with features like snapshots, backups, and replication. For extensive distributed storage needs, such as large volumes for model weights loading operations, the cluster can use an external Ceph backend. Ceph offers scalable and resilient storage with advanced features, interfaced with Kubernetes through a CSI driver and a dedicated \texttt{StorageClass}. This system ensures both virtual and HPC nodes use the appropriate storage without changing application configurations or deployment pipelines.

    \item \textbf{Examples of user-deployed services}. In a hybrid Kubernetes setup, Ollama and OpenWebUI are examples of efficient service orchestration. OpenWebUI, a frontend for inference handling on large language models, is deployed on lightweight virtual nodes suited for low resource services. Ollama, requiring GPU resources, runs on HPC nodes with NVIDIA or AMD GPUs, ensuring efficient LLM execution via Kubernetes node labels and pod affinity (e.g., \texttt{gpu=true}, \texttt{gpu.vendor=nvidia}). Cilium (based on eBPF) facilitates internal communication, optimizing network traffic between HPC nodes and control plane nodes. This case exemplifies principles like workload separation, GPU management, namespace isolation, and GitOps deployment in a scalable Kubernetes environment. Other valid examples of applications and services that users might want to deploy on such a service plane include products such as Kubeflow\footnote{\url{https://www.kubeflow.org}}, components of the Ray suite\footnote{\url{https://www.ray.io}} or on-prem deployments of tools such as Weights and Biases to cope with SaaS limits.
\end{itemize}

\subsubsection{Ongoing work and future directions} Among the elements that we are still investigating there are (1) the integration of such services with our existing IAM infrastructure, to provision access to such service plane automatically for interested projects, (2) integration with existing resource accounting and billing workflows, and (3) appropriate interfaces for the end users to foster productive adoption.

\subsubsection{Relevance} By empowering project teams to independently deploy and operate services, and offering the flexibility to adopt familiar products of their choice, we aim to enhance their productivity, enable faster responses to evolving needs in the dynamic ML landscape, and ultimately strengthen their competitiveness. Furthermore, at CSCS we intend to utilize the same service plane to deploy internally needed solutions, such as data-driven enhancements to support ticket handling.

Ensuring inference services and pipelines are portable across infrastructures helps meet future ML project needs, especially for teams with operational goals. As CSCS cannot guarantee strict SLAs for academic research, users need the flexibility to transition deployments as required. Adopting industry standards like containers and Kubernetes facilitates this process, offering smoother migration possibilities.

This proposal is relevant to address challenges \cref{item:problem_stmt:support_diverse_and_evolving_needs}, \cref{item:problem_stmt:reproducibility_and_portability} and \cref{item:problem_stmt:ops_support_constraints} as introduced in Section~\ref{sec:motivations}.

\subsection{Storage Services for ML workloads}
\label{subsec:storage}
\subsubsection{Motivation}
ML projects span several stages, including data gathering, preparation, training, inference, and sharing. Each of these stages might present distinct storage access patterns and requirements.

Data gathering typically produces many small files in heterogeneous formats. For certain use cases, such as LLM-oriented ones, preparation involves tokenizing data, usually resulting in compact representations (4–100× smaller), stored in larger files. Pre-training consumes data via random small-batch reads, thus requiring high IOPS. Such operation, generally executed in random order, presents limited caching opportunities mainly due to data being read once per training execution. Regular model checkpointing during training can generates terabytes of data across thousands of files and require medium-range persistence.

Finetuning and reinforcement learning introduce faster iteration cycles, with I/O demands similar to pre-training but at smaller scales. Inference has lighter storage needs, mostly involving model loading, and benefits from consistent, low-latency access in a service-like mode. Long-term sharing of model weights and datasets imposes archival and availability requirements.

Our current setup is based on two Lustre file systems: one SSD-backed, one HDD-backed. While Lustre excels at large sequential I/O operations, ML workloads involve frequent small, random accesses and metadata-intensive operations. On HDD-backed Lustre, seek latency can degrades performance further. Even SSD-backed Lustre performances can become impacted by metadata server load. In shared environments, resource contention further impacts performances stability. These challenges highlight the need for ML-aware data management strategies beyond traditional HPC storage configurations.

\subsubsection{Proposed Approach}
To better support scalable and efficient ML workloads on HPC infrastructure, we follow a storage strategy that integrates different architectural options, ML-friendly data management practices, and application-level software optimizations.

To fully leverage the underlying storage architecture, collaboration with users is essential. Optimizing data loading pipelines can yield significant efficiency gains, and is often done at application code level (w.r.t., data loaders). Where applicable, we promote container-native compressed formats such as SquashFS for storing static datasets, reducing I/O overhead and cold-start times in containerized environments~\cite{SquashFS}. These software-level strategies complement the physical storage design. Our approach targets the full ML life-cycle: from high-throughput pre-training to interactive fine-tuning and persistent model sharing. In this regards our approach combines a multi-tiered architecture, leveraging alternative storage options and by supporting users in application-level improvements. The key components are:

\begin{itemize}
    \item \textbf{Tiered storage architecture}: Fast-access tiers (e.g., local or remote NVMe) serve high-IOPS workloads like randomly sampled trainings, while capacity tiers (e.g., HDD-backed Lustre or archival object storage) support checkpoints, datasets, and long-term retention. Future integration of adaptive tiering based on workload profiling will enhance this, also through automation~\cite{Zhang2020TierScrubbingAA}.

    \item \textbf{NVMe and NVMe-oF}: We are investigating the usage of both node-local NVMe and fabric-attached NVMe (NVMe-oF), balancing low-latency access with operational flexibility. NVMe-oF provides near-local performance with centralized manageability, though tradeoffs include increased network contention~\cite{10579141}. A related approach is discussed under \ref{subsubsection:storage:ongoing_work}.

    \item \textbf{Object storage for unstructured data}: Ceph-like systems will be employed for scalable, metadata-rich storage of tokenized datasets, media, and model artifacts. Object storage fits well with containerized ML workflows and data sharing needs~\cite{ArsuagaRios2015}.

    \item \textbf{Lustre optimization}: While Lustre remains essential for large-scale sequential I/O, we are exploring ML-aware caching, file aggregation, and hybrid backends to mitigate its inefficiencies with small-file or metadata-intensive operations~\cite{Neuwirth2019AcceleratingNC}.

    \item \textbf{Software stack alignment}: We encourage use of container-native compressed formats (e.g., SquashFS) and user-optimized data loaders to reduce cold-start latency and I/O amplification in containerized environments~\cite{SquashFS}.
\end{itemize}

This proposed design aligns with the demands of ML workflows, offering a path to sustainable performance, resource utilization, and reproducibility in heterogeneous HPC environments.

\subsubsection{Ongoing work and future directions} \label{subsubsection:storage:ongoing_work} 
\begin{figure}[ht]
    \centering
    \includegraphics[width=\linewidth]{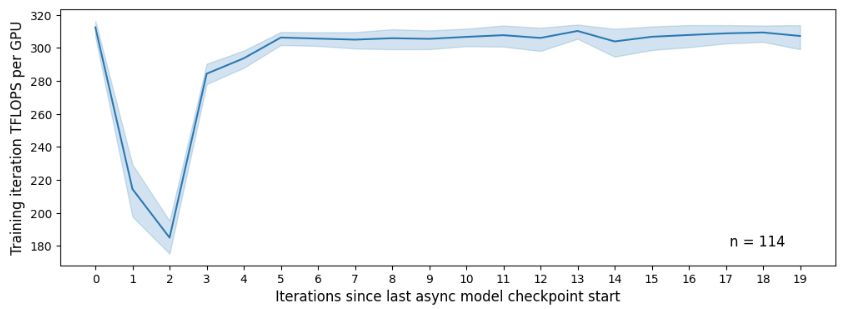}
    \caption{While beneficial to the rapid resumption of training iterations, asynchronous model checkpointing might cause consistent, temporary drops in training throughput, suggesting the need to consider diversified storage solutions tailored to varying performance demands.}
    \label{fig:storage:async_checkpointing}
    \Description{Asynchronous model checkpointing causes consistent, temporary drops in training throughput, suggesting the need to consider diversified storage solutions tailored to varying performance demands.}
\end{figure}

Current efforts focus on the following areas:

\begin{itemize}
    \item \textbf{Efficient handling of small files}: Inspired by container image storage, we support SquashFS for packing small datasets into compressed, mountable filesystems, simplifying both use and distribution. We are evaluating ComposeFS as a potential evolution of this approach.

    \item \textbf{Storage metrics integration}: As part of our observability framework (Section~\ref{subsec:observability}), we are identifying key storage metrics to integrate into our monitoring and analytics platform. This will enhance users’ ability to autonomously correlate I/O behavior with application performance and improve self-service diagnostics.

    \item \textbf{Ephemeral storage via CPU DRAM}: We are evaluating the usage of unused CPU DRAM as temporary local storage to reduce I/O performances variability on data loading phases caused by the shared nature of the current underlying storage systems.

    \item \textbf{Differentiated storage services}: Asynchronous models checkpointing~\cite{pt_async_checkpointing} is increasingly adopted by users. Although overlapped, the bursty checkpoint write operation can still affect training throughput. Distributing such operation over a longer period of time might be beneficial to the model training process \cite{llm_without_parallel_file_system}. Figure~\ref{fig:storage:async_checkpointing} illustrates an example of such situation observed on Alps, suggesting the need to consider diversified storage solutions tailored to varying performance demands. Although solutions for this can also be considered at the application level, system-level architectural options need to be considered more broadly during procurement processes.
\end{itemize}

\subsubsection{Relevance}
Combining these strategies allows supercomputing systems to balance performance, scalability, and cost while catering to both traditional HPC and modern ML workloads. An added nuance at CSCS is that we are partitioning our supercomputing infrastructure according to tenants, which needs to be supported accordingly by the underlying storage infrastructure.

The discussed approaches are relevant to address challenge \cref{item:problem_stmt:storage_offering} introduced in Section~\ref{sec:motivations}.

\subsection{Security implication of ML workloads}
\label{sec:security}
\subsubsection{Motivation}
Without IT security, a supercomputing center would quickly be exploited by malicious actors.
As a public institution having as mission to develop and operate a HPC and data research infrastructure that supports world-class science in Switzerland, we have the advantage that the goals of our user community are generally strongly aligned with our own.
The ML community security and ethical issues are not exclusive to it, or absolutely novel. Nevertheless, we want to discuss noteworthy aspects based on our experiences.

First, the ML field has high visibility due to the large societal and economic interest. This increases the stakes of all issues beyond what we commonly handle.

Second, the size and evolution speed of the community are much larger than the scientific communities that we are typically handling.
This means, for example, that codebases are a more attractive target (typical ML compute is powerful and well connected). Attacks can, for example, exploit pickle weaknesses, and hide the exploit in poisoned models\footnote{\url{https://www.darkreading.com/application-security/hugging-face-ai-platform-100-malicious-code-execution-models}}, or use dependency confusion\footnote{\url{https://www.theregister.com/2023/01/04/pypi_pytorch_dependency_attack}}. We encountered attempts to run a captcha cracker on our infrastructure.

Finally, generally we try to be a neutral infrastructure provider, and externalize the ethical considerations to other institutions and the peer review process. This is  not fully possible, especially when the safety of infrastructure is at stake.

ML data gathering collects large datasets on which the ML models are then trained. The origin of these datasets depends on the field. Public data downloaded from the Internet is often a component of these datasets. AI crawlers, which are causing increasing hosting costs for content providers\footnote{\url{https://arstechnica.com/information-technology/2025/04/ai-bots-strain-wikimedia-as-bandwidth-surges-50}}, are increasingly considered unfairly profiting (or even stealing) while giving little back in return. Entities displeased by this situation developed tools like Nepenthes\footnote{\url{https://zadzmo.org/code/nepenthes}} and iocaine\footnote{\url{https://crates.io/crates/iocaine}}, traps aimed at slowing down these web crawlers.

Some common datasets used to train models, such as the coyo-dataset, cannot be easily downloaded anymore because many of the images are now missing\footnote{\url{https://github.com/kakaobrain/coyo-dataset/tree/main/download##missing-images}}. This situation is made worse by some of these pointers (URLs) which are not just dangling, but replaced with malware contents. Downloading contents from these affected URLs might lead to being gray- or black-listed by automatic protection services provided by the large content distribution networks, which are motivated by financial interests.

Sharing the datasets (which is, in part, one of the goals that the Swiss AI Initiative has on our infrastructure) also aims at fostering trust on the models trained on them. In practice, this is rarely done because hosting data carries more legal liability than hosting data references. Also, hosting data requires methods to identify and remove problematic data, adding complexity to the hosting party and making future reproducibility harder to maintain.

\subsubsection{Proposed Approach}
Security in HPC environments has no silver bullet; it requires continuous monitoring, timely responses, and proactive user engagement. Our proposed approach includes:

\begin{itemize}
    \item \textbf{Centralized container images management:} We encourage the use of container registries that provide automated vulnerabilities scanning, and enhances control over the software supply chain.
    
    \item \textbf{Network activity oversight:} We aim at continuous monitoring of outbound internet traffic and ensure our ability to promptly revoke access from specific nodes or jobs in response to suspicious behavior.
    
    \item \textbf{User awareness and training:} Promote security hygiene by educating users on the risks of running unverified code or models, particularly those sourced from untrusted or opaque origins.
    
    \item \textbf{Internet access policy enforcement:} Web crawling activities, e.g., for the purpose of building raw datasets from web content, is not allowed on our Alps infrastructure. Internet access is available, but bulk download from websites that are not aware and have not agreed to receive our high-load Internet traffic is against our policies. For this reason, we have rules configured to continuously monitor the volumes of Internet traffic. Kubernetes-based options, outside of the Alps infrastructure, operate under slightly more flexible policies, but still require vigilant oversight. As an HPC center, we risk having our public IPs blacklisted by external service providers due to problematic (though well-intentioned) activities performed by our users on our systems.
\end{itemize}

\subsubsection{Design elements}
Our security design emphasizes proactive communication, real-time monitoring, and the promotion of trusted software options. Effective communication channels with users, such as comprehensive documentation, a dedicated Slack channel, and weekly drop-in sessions, are fundamental to keep the awareness high regarding security risks and to enable rapid response to incidents. At the network level, a firewall monitors all outbound traffic, flags suspicious activity, and, when necessary, blocks sources to initiate further investigation. We recommend using container registry services with security features like vulnerability scans of images and dependencies to secure the software supply chains.

\subsubsection{Relevance}
Security and infrastructure integrity are foundational prerequisites for delivering any computational service, including those supporting ML and scientific research. Without robust safeguards, malicious activity can compromise resource availability, data integrity, user trust, and the reputation of the institutions involved. As such, our ability to provide reliable HPC capabilities is critically dependent on proactive security measures, responsible data governance, and community-wide awareness, especially as we support a growing and dynamic ML user base.


\section{Discussion}
\label{sec:discussion}
The aim of the preliminary work presented in this paper is to address the delicate tension between the urgency to deliver immediately usable solutions to ML users on Alps, and the necessity of following a principled systematic approach to analyze user needs and select architectural options.

The fast-evolving nature of ML, characterized by frequent introduction of new algorithms, methods and services, together with their rapid commoditization by vendors, makes it difficult to forecast the lasting impact of any single effort. As a result, time-to-market or time-to-publish becomes a critical objective at the expense of mid and long development planning and architectural rigor. Nevertheless, the deliberate exposure of ML communities to HPC services aims to leverage collective insight and foster converged AI and HPC infrastructure services.

A further challenge lies in the diversity of infrastructure and service expectations by the different groups in the ML communities. Although performance and GPU access are shared priorities, user preferences vary greatly, from low-level control to SaaS-like abstractions. Those expectations are reflected in the different elements presented in this work such as spanning container-based environments, service planes, early node vetting, or data access performance. However, providing these services demands additional effort, attention, and resources to keep up with the developments in the field of ML, posing challenges for publicly funded centers like ours.

Many of the proposed components reflect a broader design pattern: operationalizing expertise related to large-scale ML on HPC infrastructure. Tools such as node vetting, observability dashboards, and GPU saturation scorers encapsulate best practices into infrastructure usability, reducing the barrier of entry for less-experienced users and promoting a culture of shared expertise among the HPC providers and the ML users.

In a similar vein, our evolving storage architecture underscores the need to reimagine data services for ML workloads, which span distinct phases: data acquisition, preprocessing, training, inference, and publicizing, each with unique I/O patterns. Traditional parallel file systems are suboptimal for the fine-grained and bursty operations typical in ML. The integration of tiered storage, NVMe-over-Fabrics, and compressed file systems (e.g., \texttt{squashfs}) marks a promising step, but architectural implications warrant further investigation.

While HPC infrastructure is naturally best suited for large-scale training jobs, arguably the HPC mandate, it is clear that the full lifecycle of ML research involves diverse, smaller-scale tasks. These include project development iterations, dataset preparation, and inference workloads. While we aim to support a broad range of needs, our mission focuses on supporting, with our specialized infrastructure, world-class science requiring large-scale workloads. This principle can serve as a compass when navigating trade-offs, ensuring that our efforts remain aligned with our institutional objectives.

Furthermore, a center such as CSCS has significant experience in operationalizing computational workloads, such as running national weather forecasts \cite{maurocug25}, an expertise that could be essential to extract societal value from academic ML research.

Ultimately, while this work proposes concrete technical responses to the identified gaps, it also acknowledges the absence of a comprehensive framework for systematically evaluating ML user needs and for rigorously evaluating solution options. By sharing our initial design trajectory, we seek feedback and alignment with peer institutions navigating comparable challenges in enabling ML workloads on HPC systems.

\section{Conclusions}
\label{sec:conclusions}
The proposals described in this work represent a concrete response to the changing expectations placed on HPC institutions like CSCS. While each of the components was motivated by observed needs of the ML community, this effort remains an ongoing exploration and is not a final answer. It is rather a set of directional steps, subject to change as requirements deepen and the ML landscape continues to evolve.

Our trajectory aligns with a broader, observable trend: HPC facilities around the world (and our vendors alike) are increasingly pulled towards the needs of AI/ML workloads. This is reflected in initiatives such as EuroHPC's AI Factories and national strategies that frame AI as a key area of competitiveness not just in academia. We believe that HPC centers are uniquely positioned to enable these goals, not only by providing computational power but also by offering a high-quality operational environment for productive and scalable ML workloads.

\begin{acks}
We acknowledge the contributions of our user community, particularly members of the Swiss AI Initiative, whose collaboration and feedback are instrumental in shaping the design of a robust and innovative ML platform.

We also acknowledge the usage of tools such as Writefull, ChatGPT, and Gemini, which we used to improve the clarity and readability of texts throughout this document. All generated suggestions were manually reviewed and/or edited before adoption to ensure fidelity to the authors' original intent. These tools were used strictly for editing purposes and not for generating ideas or data.

\end{acks}


\printbibliography

\end{document}